\documentclass[prd,aps,floats,floatfix,eqsecnum,nofootinbib]{revtex4}
\usepackage{graphicx,amssymb,amsbsy,bm,amsmath,rotating}
\usepackage{psfrag}
\newcommand{\avg}[1]{\langle #1 \rangle}

\newcommand{\be}{\begin{equation}}
\newcommand{\ee}{\end{equation}}
\newcommand{\bea}{\begin{eqnarray}}
\newcommand{\eea}{\end{eqnarray}}

\newcommand{\vx}{\ensuremath{\vec{x}}}

\newcommand{\vq}{\ensuremath{\vec{q}}}

\newcommand{\rv}{\ensuremath{\vec{r}}}
\newcommand{\pv}{\ensuremath{\vec{p}}}
\begin{document}
\title{Constant surface gravity and density profile of dark matter}
\author{H. J. de Vega}
\email{devega@lpthe.jussieu.fr} \affiliation{
LPTHE, Laboratoire Associ\'e au CNRS UMR 7589,\\
Universit\'e Pierre et Marie Curie (Paris VI) et Denis Diderot
(Paris VII), \\ Tour 24, 5 \`eme. \'etage, 4, Place Jussieu,
75252 Paris, Cedex 05, France and 
 Observatoire de Paris, LERMA, Laboratoire Associ\'e au
CNRS UMR 8112, \\61, Avenue de l'Observatoire, 75014 Paris, France.}
\author{N. G. Sanchez}
\email{Norma.Sanchez@obspm.fr}
\affiliation{Observatoire de Paris, LERMA, Laboratoire Associ\'e au
CNRS UMR 8112, \\61, Avenue de l'Observatoire, 75014 Paris, France.}

\begin{abstract}
Cumulative observational evidence confirm that the surface 
gravity of dark matter (DM) halos $ \mu_{0 D} = r_0 \; \rho_0 $ where $ r_0 \;$ and $ \rho_0 $ 
are the halo core radius and central density, respectively, is nearly {\bf constant} 
and independent of galaxy luminosity for a high number of galactic systems 
(spirals, dwarf irregular and spheroidals, elliptics) 
spanning over $14$ magnitudes in luminosity and of different Hubble types. 
Remarkably, its numerical value $ \mu_{0 D} \simeq 140 \; M_{\odot}/{\rm pc}^2 =  
(18.6 \; {\rm Mev})^3 $ is approximately {\bf the same} (up to a factor of two) in all 
these systems. First, we present the physical consequences 
of the independence of $ \mu_{0  D} $ on $ r_0 $: the energy scales as the volume  $ \sim r_0^3 $ 
while the mass and the entropy scale as the surface $ \sim r_0^2 $
and the surface times $ \log r_0$, respectively. Namely, the entropy scales similarly to
 the black-hole entropy but with a much smaller coefficient. Second,
we compute  the surface gravity and the density profile for small scales from first principles
and the evolution of primordial density fluctuations since the end of inflation till today
using the linearized Boltzmann-Vlasov equation. The 
density profile $ \rho_{lin}(r) $ obtained in this way decreases as $ r^{-1-n_s/2} $ 
for intermediate scales where $ n_s\simeq 0.964 $ is the {\bf primordial} spectral index. 
This scaling is in remarkable agreement with the empirical behaviour 
found observationally and in $N$-body simulations:
$ r^{-1.6\pm 0.4} $. The observed value of $ \mu_{0  D} $ indicates that the DM 
particle mass $ m $ is in the keV scale. 
The theoretically derived density profiles  $ \rho_{lin}(r) $ turn to 
be {\bf cored} for $ m $ in the keV scale and they look as {\bf cusped} for
$ m $ in the GeV scale or beyond. 
We consider both fermions and bosons as DM particles decoupling either ultrarelativistically or 
non-relativistically. 
Our results do {\bf not} use any particle physics 
model and vary slightly with the statistics of the DM particle. 
\end{abstract}
\maketitle
\tableofcontents

\section{Observational evidences}

Growing recent findings point towards a constant dark matter (DM) surface gravity 
$ \mu_{0 D} $ in galaxy DM halos \citep{kor,dona,span}. Namely, the product $ \mu_{0  D}
\equiv r_0 \; \rho_0 $ where  $ r_0 $ and $ \rho_0 $ are the halo core radius and central
density, respectively, is nearly {\bf constant}, over a large number of galaxies of different kinds
\be\label{valmu}
\mu_{0  D} \simeq 140 \; \frac{M_{\odot}}{{\rm pc}^2} = 6400 \; 
{\rm MeV}^3 = (18.6 \; {\rm Mev})^3 \; , 
\ee 
while $ r_0 $ varies by two orders of magnitude \citep{kor,dona,span}: 
\be\label{datos}
0.3 \,  {\rm kpc} < r_0 < 30  \;  {\rm kpc} \quad   {\rm and}   \quad
 10^{-25}  \; {\rm g/ cm^3} \leq \rho_0 \leq 6 \times 10^{-23} \; 
{\rm g/ cm^3} \; . 
\ee
This finding relates to data sets (high quality rotation curves, kinematics, galaxy-galaxy weak 
lensing signals) for many galactic systems spanning over 14 magnitudes in luminosity and of 
different Hubble type, dwarf disk and spheroidals, spirals, ellipticals. 
In spite of their different properties, $ \mu_{0 D} $ in galaxies is essentially 
{\bf independent} of their luminosity and mass. The surface gravity $ \mu_{0 D} $ is also the 
surface density.

For luminous matter, the surface gravity takes also the value eq.(\ref{valmu})
provided $ \mu_{0 D} $ is obtained as the product of the halo core radius  $ r_0 $  times the
density at $ r_0 $ \citep{gg}.

\medskip

It must be noticed that relations analogous to the eq.(\ref{valmu}) are also
known for interstellar molecular clouds in our Galaxy \citep{larson}. 
One of the scaling laws put forward by Larson \citep{larson} 
states that the surface gravity (column density) $ \mu_{0 D} $ is approximately a 
{\bf constant} over more than four orders of magnitude of scales 
$ 0.001 \, {\rm pc} < r_0 < 100 $ pc. The values given in \citet{larson} are:
$$
\mu_{0 D} = 10.5 \; 10^{21}\; \frac{m_{H_2}}{{\rm cm}^2} = 162 \; \frac{M_{\odot}}{{\rm pc}^2}
$$
where $ m_{H_2} $ stands for the mass of the Hydrogen molecule, main constituent of the 
interstellar clouds. Recent data averaged over high density regions of Taurus give \citep{gold}
\be\label{nube}
\mu_{0 D} = 5.14 \; 10^{21}\; \frac{m_{H_2}}{{\rm cm}^2} = 80  \; \frac{M_{\odot}}{{\rm pc}^2}
\ee
The mean density of structures in the ISM vary between $ 10 $ and $ 10^5 $ atoms/cm${}^3$,
significantly above the mean ISM density which is about $ 0.1 $ atoms/cm${}^3$ or
$ 1.6 \; 10^{-25} $ g/cm${}^3$. Hence eqs.(\ref{valmu}) and (\ref{nube}) are
verified both for molecular clouds and galaxies (up to a factor $ 2 $). 

\medskip

The quantities $ r_0 $ and $ \rho_0 $ depend on the particular galaxy 
(or molecular cloud) chosen and are therefore functions of the past history 
of the galaxy (or cloud). Instead, the product $ \mu_{0  D} = r_0 \; \rho_0 $
given by eq.(\ref{valmu}) is an {\bf universal} number for all galactic 
systems and molecular clouds and hence independent of the previous history 
of the system.  Therefore, $ \mu_{0  D} $ can only depend on 
{\bf universal quantities}. Since $ \mu_{0  D} $ is the same 
(up to a factor two) for molecular clouds and galaxies, the action of
self-gravity (both of baryonic and dark matter) should be responsable of its 
value since it is the only common physical mechanism to all these objects. 
Indeed, other processes play a role in the physics
of molecular clouds and galaxies and can affect the surface gravity
deviating it from the universal value eq.(\ref{valmu}) by a factor of two or so
(see for example \citet{Hetal}).
These processes are therefore {\bf subdominant} with respect to self-gravity.
For example, the observed mean surface gravity in the M64 galaxy is $ 10^{22} \; 
m_{H_2}/cm^2 $ \citep{RB} within 15 \% of our equation (\ref{valmu}).

\medskip

As stressed in \citet{disvdb,disvdb2,disvdb3,disvdb4}, a single parameter should control
the galaxy structure implying that functional relations must constrain 
galaxy parameters as mass, size, baryon-fraction, etc. We propose that
the surface gravity $ \mu_{0  D} $ (as a function of these galaxy parameters)
can be one of these functional relations necessary to explain the parameter correlations 
presented in \citet{disvdb,disvdb2,disvdb3,disvdb4}. 

This implies that $ \mu_{0  D} $ is independent of the baryon-fraction value.
Such independence is consistent with the fact that 
molecular clouds (dominated by baryons) have similar $ \mu_{0  D} $ that DM dominated galaxies
[see eqs.(\ref{nube}) and (\ref{valmu})].

\medskip

We analyze in the next section how the mass, the energy and the entropy
scale with the size $ r_0 $ as a consequence that $ \mu_{0  D} $ is a universal
constant and therefore independent of $ r_0 $ in the context of kinetic theory
for self-gravitating systems.

In section \ref{perfBV} we derive the value of $ \mu_{0  D} $ 
and the density profile for small scales from first principles
We use as appropriate initial conditions the primordial inflationary power spectrum and we follow 
the evolution through the radiation and matter dominated eras using the 
linearized Boltzmann-Vlasov equation for self-gravitating DM. In sections \ref{pld} and 
\ref{conclu} we derive
the properties and implications of the obtained linear density profiles and surface gravity. 
The derivations presented in sections \ref{perfBV} to \ref{conclu} do not rely on the analysis 
made in sec. \ref{exten} as they are independent of it. 

\section{Constant surface gravity and the scaling of the mass, 
energy and entropy}\label{exten}

Considering that the dark matter distribution in galaxies is characterized by a scale
$ r_0 $, the matter density can be written as
\be\label{dens}
\rho(r) =  \rho_0 \; F\left(\frac{r}{r_0}\right) \quad , \quad  F(0) = 1  \; .
\ee
Algebraic fits to the DM cored density profile \citep{span,bur} and thermal 
profiles are particular examples of eq.(\ref{dens}). We have for the Burkert \citep{bur} and
Spano \citep{span} profiles (denoted $ F_{B} $ and $ F_{S} $, respectively):
$$
F_{B}(x) =  \frac1{(1+x) \; (1 + x^2)} \; , \; F_{S}(x) =  
\frac1{\left(1 + x^2\right)^\frac32} \; , \; x \equiv \frac{r}{r_0} \; ,
$$
Notice that both the Burkert and the Spano profiles decay for large distances with
the same $ 1/r^3 $ tail as the cuspy Navarro-Frenk-White profile.

Each galaxy can be considered as an isolated system.
The virial theorem for isolated self-gravitating systems (that is, zero external pressure)
states that the total energy $ E $ is 
related to the average potential energy $ \avg U $ and the average kinetic energy $ \avg K $  
by \citep{ll}
\be\label{vir}
E = \frac12 \; {\avg U} = -{\avg K} \; .
\ee
We can therefore express the total energy $ E $ in terms of the average gravitational
potential energy as
\be\label{estim}
E = - \frac14 \; G \; \int \frac{d^3r \; d^3r'}{|\rv-\rv'|} \; \avg{\rho(r) \; \rho(r')}=
- \frac14 \; G \; \rho_0^2 \; r_0^5 \int \frac{d^3x \; d^3x'}{|\vx-\vx'|} \; 
\avg{F(x) \; F(x')} \; .
\ee
Hence, since the integrals over $ \vx $ and $ \vx' $ in eq.(\ref{estim}) are of order one,
the energy divided by the characteristic volume $ r_0^3 $ goes as
\be\label{ext}
\frac{-E}{r_0^3} \sim G \; \rho_0^2 \; r_0^2 = G \; \mu_{0  D}^2 \; .
\ee
The mass density eq.(\ref{dens}) inserted in the Poisson equation
$$
\nabla^2 \phi(r) = 4 \, \pi \; G \; \rho(r) \; ,
$$
yields a gravitational potential $ \phi(r) $ of the form
\be\label{potg}
\phi(r) = G \; r_0^2 \; \rho_0 \; \Phi(x) \; ,
\ee
where $ \Phi(x) $ is a dimensionless function.

The matter density $ \rho(r) $ can be expressed in the kinetic theory framework as
an integral over the velocities 
$$
\rho(r) = m \; \int f(\pv,\rv) \; d^3 p
$$
where $ f(\pv,\rv) $ is the distribution function and $ m $ the mass
of the DM particles. $ f(\pv,\rv) $ obeys the Boltzmann-Vlasov equation 
\be\label{ebv}
\frac{\partial f}{\partial t} + \frac1{m} \; \pv \cdot \frac{\partial f}{\partial \rv} 
- m \; \frac{\partial \phi}{\partial \rv}  \cdot  \frac{\partial f}{\partial \pv}  = 0
\ee
and it is normalized by the total number of particles $ N $ as
\be\label{N}
\int f(\pv,\rv) \; d^3 p \;  d^3 r = N \; .
\ee
The appropriate dimensionless variables for the Boltzmann-Vlasov equation (\ref{ebv})
and the gravitational potential eq.(\ref{potg}) are defined as
\be\label{trafv}
\rv = r_0 \; \vx \quad , \quad \pv =  m \; r_0 \; \sqrt{G \; \rho_0} \; \vq \quad , \quad
t = \frac{\tau}{\sqrt{G \; \rho_0}} \; ,
\ee
Here $ \vq $ and $ \tau $ stand for the dimensionless momentum and time, respectively.
It is convenient to introduce a dimensionless distribution function
\be\label{trafof}
{\cal F} (\vq,\vx) =  m^4 \; r_0^3 \; G^{\frac32} \; \sqrt{\rho_0} \;  f(\pv,\rv) \; ,
\ee
which enjoys the property,
$$
\int d^3 q \; {\cal F} (\vq,\vx) = F(x) \quad .
$$
Since the integral of  $ F(x) $ over a volume of order one in $ \vx $ is also of order one,
the total mass from eq.(\ref{dens}) scales as 
\be\label{MN}
M = m \; N \sim  r_0^3 \; \rho_0 = \mu_{0 D} \; r_0^2 \; .
\ee
and
\be\label{norF}
\int d^3 q \; d^3x \;  {\cal F} (\vq,\vx) = {\cal O}(1)
\ee
where $ {\cal O}(1) $ means  $ {\cal O}([r_0]^0) $, independent of the halo size $ r_0 $. 

We can estimate the entropy 
$$
S = \int f(\pv,\rv) \; \log f(\pv,\rv) \; d^3 p \;  d^3 r \; .
$$
From eqs. (\ref{trafv}), (\ref{trafof}) and (\ref{norF}) we obtain
\be\label{entro}
S \sim r_0^3 \; \frac{\rho_0}{m} \; \log  r_0 = r_0^2 \; \log r_0  \; \frac{\mu_{0 D}}{m} \; .
\ee
The average kinetic energy follows from the distribution function eq.(\ref{trafof}) to be
\be\label{K}
K = \frac1{2\, m} \; \int f(\pv,\rv) \; \pv^{\, 2} \; d^3 p \;  d^3 r 
= \frac12 \;G \; r_0^5 \; \rho_0^2 \int d^3 q \; d^3x \;  \vq^{\, 2} \; {\cal F} (\vq,\vx) 
\sim G \; r_0^5 \; \rho_0^2 = G \; \mu_{0  D}^2 \;  r_0^3 \; ,
\ee
and similarly for the total energy eq.(\ref{ext}).

The average squared velocity 
$$ 
\avg{v^2} = \frac{\avg{\pv^{\, 2}}}{m^2} 
$$
follows from eq.(\ref{trafv}) to be equal to
\be\label{v}
\avg{v^2} = \frac1{m^2} \; 
\frac{\int \pv^{\, 2} \; f(\pv,\rv) \; d^3 p \;  d^3 r}{\int f(\pv,\rv) \; d^3 p \;  d^3 r}
\sim  G \; \mu_{0  D} \; r_0 \; .
\ee
Notice that although the above derivation applies to the DM mass distribution
the results may be also true for systems where the baryonic mass is important
and hence for such systems
the above derivation should be generalized adding the baryonic contribution.

\medskip

We thus find that a constant surface gravity $ \mu_{0  D} $ (that is, independent of
the halo radius $ r_0 $) implies that the energy (total, potential
and kinetic) scales as the volume ($ \sim r_0^3 $) eqs.(\ref{estim}), (\ref{ext}) and (\ref{K})
 while the total mass and entropy scale as the surface 
($ \sim r_0^2 $) and the surface times $ \log r_0  $, respectively [eqs. (\ref{MN}) and (\ref{entro})].

This scaling follows from the long range nature of the gravitational
interactions plus the fact that this system is not in thermal equilibrium
but in quasi-equilibrium configurations.

The {\bf entropy} scales as the {\bf surface} also for black-holes.
However, for black-holes of mass $ M $ and area $ A = 16 \, \pi \; G^2 \; M^2 $,
the entropy $ S_{BH} = A /(4 \; G) = 4 \, \pi \; G \; M^2 $.
That is, the proportionality coefficients $ c $ between entropy and area are
 very different:
$$
c_{gal}=\frac{S_{gal}}{r_0^2} \sim \frac{\mu_{0 D}}{m} \quad , \quad
c_{BH}= \frac{S_{BH}}{A} = \frac1{4 \; G} \quad {\rm which ~ implies} \quad
\frac{c_{BH}}{c_{gal}} \sim \frac{m}{\rm keV} \; 10^{36}
$$
showing that the entropy per unit area of the galaxy is much smaller 
than the entropy of a black-hole. In other words, the Bekenstein 
bound for the entropy of physical is well satisfied here.

Notice that the surface gravity acceleration is given by $ G \; \mu_{0  D} $.

\medskip

We derive in the next section $ \mu_{0  D} $ as a dynamical scale 
determined by gravitational clustering. We consider in what follows DM
in galaxies.

\section{The density profile and the 
 surface gravity from the linearized Boltzmann-Vlasov equation}\label{perfBV}

The mass density $ \rho_{lin}(r) $ can be evaluated theoretically solving the 
linearized Boltzmann-Vlasov equation for self-gravitating DM in the matter 
dominated (MD) era. It is convenient to recast such 
equation as an integral equation, namely the Gilbert equation which is a 
Volterra equation of second kind \citep{gbs,gbs2,nos2}. To linear order in 
perturbations the distribution function of the decoupled particles can be 
written as
$$
f(\vec{x},\vec{p};t) = g \; f_0(p)+ F_1(\vec{x},\vec{p};t) 
$$
where $ \vec{x},\vec{p} $ are comoving coordinates, $ g $ 
is the number of internal degrees of freedom 
of the DM particle, typically $ 1 \leq g \leq 4 $, $ f_0(p) $ is the thermal equilibrium 
unperturbed distribution function at the decoupling temperature $ T_d $ normalized by
\be\label{norf}
m \; g \int \frac{d^3p}{(2\pi)^3} \;  f_0(p)  = \rho_{DM} = \Omega_M \; \rho_c \; ,
\ee
where $ \Omega_M =  0.233 $ is the DM fraction, $ \rho_c $ is the critical density of the Universe
\be\label{roc}
\rho_c = 3 \, M_{Pl}^2 \; H_0^2 = (2.518 \; {\rm meV})^4 \quad , \quad 1 \, 
{\rm meV} = 10^{-3} \, {\rm eV}
\ee
and $ H_0 = 1.5028 \; 10^{-42} $ GeV.
Terms of order higher than one in $ F_1 $ are neglected in the Boltzmann-Vlasov equation.
The physical initial conditions at $ t_{eq} $, the beginning of the MD era, are 
\be\label{fini}
f(\vec{x},\vec{p};t_{eq}) = g \; f_0(p)[1+ \delta(\vec{x},t_{eq})] \quad ,  \quad
F_1(\vec{x},\vec{p};t_{eq}) = g \; f_0(p) \; \delta(\vec{x},t_{eq}) \; ,
\ee
where $ \delta(\vec{x},t_{eq}) $ are the density fluctuations by the beginning of the MD era.

It is useful to Fourier transform over $ \vec{x} $ and 
integrate the fluctuations $ F_1(\vec{x},\vec{p};t) $ over the momentum $ \vec{p} $,
\be\label{dgra}
\Delta(k,t) \equiv m \; \int \frac{d^3p}{(2\pi)^3} \; F_1( \vec{k}, \vec{p}\,;t) \quad
{\rm where} \quad F_1( \vec{k}, \vec{p}\,;t) = \int d^3x \; e^{-i \, \vec{x} \cdot \vec{k}}
\; F_1(\vec{x},\vec{p};t) \; .
\ee
Its Fourier transform provides the matter density fluctuations $ \rho_{lin}(r) $ in linear
approximation today
\be\label{defro}
\rho_{lin}(r) = \frac1{2 \, \pi^2 \; r} \; \int_0^{\infty} k \; dk \; \sin(k \, r) \;  
\Delta(k,t_{\rm today}) \; ,
\ee
where as customary we considered a spherical symmetric distribution. 

We therefore have as initial conditions using eqs.(\ref{fini}) and (\ref{dgra}),
\be\label{Delteq}
\Delta(k,t_{eq})=\Omega_M \; \rho_c \; \delta(k,t_{eq}) \; ,
\ee
The present linear treatment is valid for small scales $ k > k_{eq} $. Non-linear effects
become important for large scales $ k < k_{eq} $ and call for the use of the full (non-linear) 
Boltzmann-Vlasov equation or $N$-body simulations.

\medskip

The density fluctuations $ \delta(k,t) $ by the end of the radiation dominated 
(RD) era can be obtained analytically for
subhorizon wavenumbers \citep{dod,husu}. The initial conditions for $ \delta(k,t) $ in
the RD era are the primordial inflationary fluctuations. 
With such initial conditions and solving the 
fluid equations for DM during the RD era yields \citep{dod,husu}
\be\label{dini}
\delta(k,t_{eq}) = \frac12 \; A \; |\phi_k| \; \left\{ 5 \; \log\left[\frac{4 \, \sqrt2 \; 
B \; k}{k_{eq}\; (1 + \sqrt2 )^2 }\right] + 6 \, \sqrt2 -15 \right\} \; \sqrt{V} 
= \frac52 \; A \; |\phi_k| \;
\log\left(0.2637 \; B \; \frac{k}{k_{eq}} \right) \; \sqrt{V}\; ,
\ee
where $ V = b_1/k_{eq}^3 $ with $ b_1 \sim 1 $
is the comoving horizon volume by equilibration. Namely,
all fluctuations with $ k > k_{eq} $ that were inside the horizon by equilibration
are relevant here. More explicitly, 
$ k_{eq} \simeq 42.04 \; H_0 = 9.88 \; {\rm Gpc}^{-1} $ \citep{dod} and
\be\label{V}
\sqrt{V} \simeq \frac {b_1}{k_{eq}^\frac32}  \simeq \frac{ b_1 \; b_0}{H_0^\frac32} 
\quad {\rm where} \quad b_0 \simeq 3.669 \; 10^{-3} \; .
\ee
$ A \simeq 9.6 $ and $ B \simeq 0.44 $ are constants that follow evolving
the fluid equations \citep{husu}, $ \phi_k $ are the  primordial inflationary fluctuations of the 
newtonian potential \citep{biblia,dod}
\be\label{fikp}
|\phi_k| = \sqrt2 \; \pi \; \frac{| \Delta_0 |}{k^\frac32} \; 
\left(\frac{k}{k_0}\right)^{\frac{n_s-1}2} \; ,
\ee
$ | \Delta_0 | $ stands for the primordial power amplitude, $ n_s $ is the spectral index,
and $ k_0 $  the pivot wavenumber \citep{WMAP,biblia},
\be\label{ns}
|\Delta_0 | \simeq 4.94 \; 10^{-5} \quad , \quad n_s \simeq 0.964 \quad , \quad
k_0 = 2 \; {\rm Gpc}^{-1} \; .
\ee
It is convenient to define
\be\label{dchi}
{\widehat \Delta}(k,t) \equiv \frac{\Delta(k,t)}{\Delta(k,t_{eq})}
\ee
Then, the Gilbert equation takes the form \citep{nos2}
\be \label{gil2}
{\widehat \Delta}(k,u) - \frac6{\alpha}   \int^u_0 \Pi[\alpha \; (u-u')] \;
\frac{{\widehat \Delta}(k,u')}{[1-u']^2} \; du' = I[\alpha \, u]   \; , 
\ee
where,
\bea
\Pi[z] &=& \frac1{I_2} \; \int_0^\infty dy \; y \; f_0(y) \; \sin(y \, z)   \quad ,  \quad
I[z] = \frac1{I_2} \; \int_0^\infty dy \; y \; f_0(y) \; \frac{\sin(y \, z)}{z}  \quad ,  \quad
I_2=\int_0^\infty dy \; y^2 \; f_0(y)  \;  , \cr \cr
y &\equiv& \frac{p}{T_d}  \quad ,  \quad
 z \equiv \alpha \; u \quad , \quad \alpha\equiv \frac{2\, k}{H_0} \; 
\sqrt{\frac{1 + z_{eq}}{\Omega_M}} \; \frac{T_d}{m} \quad ,  \quad
1+z_{eq} = \frac1{a_{eq}} \simeq 3200 \; .
\eea
$ u $ is a dimensionless time variable related to the scale factor by
$$
u = 1 - \sqrt{\frac{a_{eq}}{a}} \quad , \quad a(u) = \frac{a_{eq}}{(1-u)^2} \quad , 
$$
$$
0 \leq u \leq  u_{\rm today} = 1- \sqrt{a_{eq}} \simeq 0.982 \quad , \quad a({\rm today}) = 1 \; .
$$
It follows from the resolution of the Gilbert equation eq.(\ref{gil2}) that for late times
the solution grows as \citep{nos2}
\be\label{dasy}
{\widehat \Delta}(k,t) \buildrel{t \to t_{\rm today}}\over= \frac35 \; T(k) \; (1 + z_{eq})
\ee
where $ T(k) $ stands for the transfer function. That is, 
$ {\widehat \Delta}(k,t) $ grows proportional to the scale factor in the linear approximation 
for all $ k < k_{fs} $. The free-streaming
comoving wavenumber $ k_{fs} $ increases with time as $ 1/\sqrt{1+z} $. 

\begin{figure}[h]
\begin{turn}{-90}
\psfrag{"tkfdgiR.dat"}{Fermi-Dirac}
\psfrag{"tkbegiR.dat"}{Bose-Einstein}
\psfrag{"r2tkmbgiR.dat"}{Maxwell-Boltzmann}
\includegraphics[height=17cm,width=10cm]{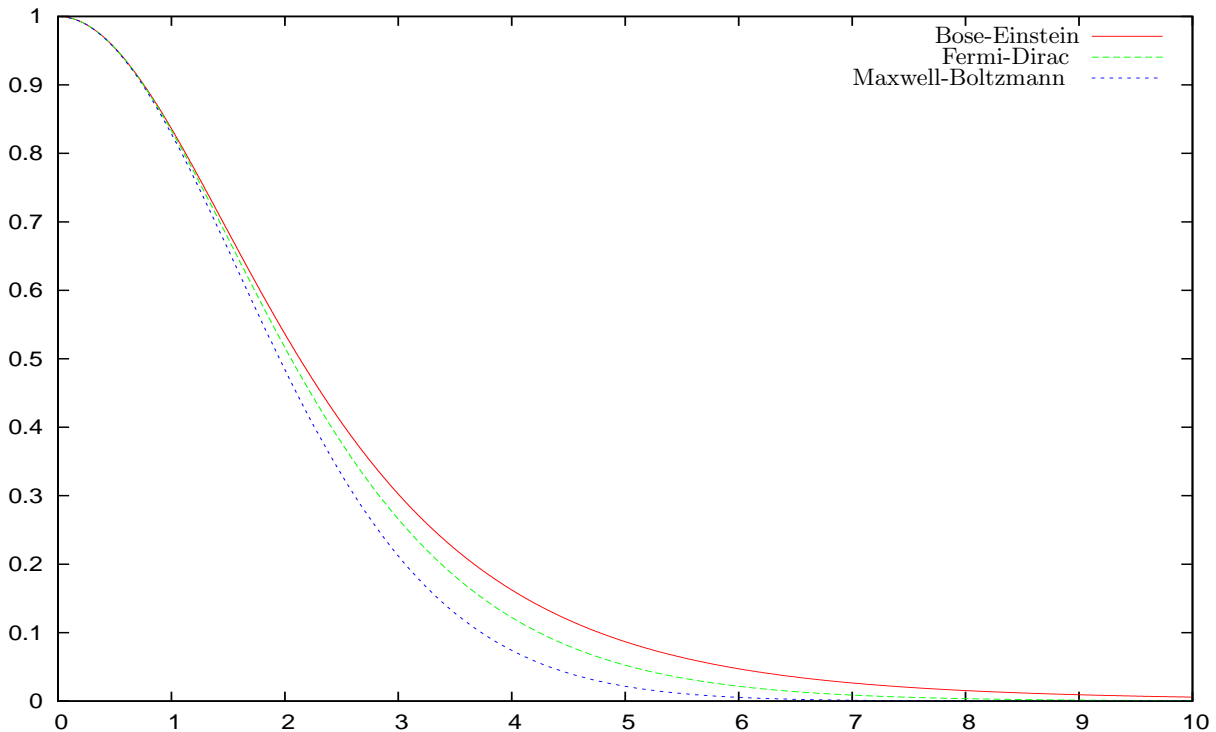}
\end{turn}
\caption{The transfer function $ T(k) $ vs. $ \gamma = k \; r_{lin} $
for Fermions and Bosons decoupling ultrarelativistically and for particles decoupling 
non-relativistically (Maxwell-Boltzmann statistics). We see that $ T(k) $ decays for increasing
$ k $ with a characteristic scale $ \sim  1/r_{lin} \sim k_{fs} $ [see eq.(\ref{defga})].}
\label{tk}
\end{figure}

\medskip  

$ T(k) $ is obtained by solving the Gilbert equation (\ref{gil2}) \citep{nos2}. 
We plot in fig. \ref{tk} $ T(k) $ for Fermions (FD) and Bosons (BE) decoupling ultrarelativistically 
and for particles decoupling non-relativistically (Maxwell-Boltzmann statistics, MB).
$ T(k) $ enjoys the properties $ T(0) = 1 $ and $ T(k \to \infty) = 0 . 
\; T(k) $ decreases with $ k $ according to the characteristic scale given by the free streaming 
wavenumber $ k_{fs} $ where $ l_{fs} = \sqrt6/k_{fs} $ is the free streaming length \citep{nos2}. 
$ T(k) $ shows little variation with the statistics of the DM particle.
The explicit expression of the comoving free streaming length is 
\be\label{lfs}
l_{fs} =\frac{2\, \sqrt3}{H_0} \; \sigma_{DM} \; \sqrt{\frac{1 + z_{eq}}{\Omega_M}} \quad , \quad
\sigma^2_{DM} \equiv \frac13 \; <v^2> \; .
\ee
$ \sigma_{DM} $ stands for the primordial comoving squared velocity 
dispersion of the DM particles. That is, the velocity dispersion computed from the thermal equilibrium 
distribution function $ f_0(p) $ which can be expressed as
\be\label{sigma}
\sigma_{DM} = \sqrt{\frac{I_4}{3 \, I_2}} \; \frac{T_d}{m} \quad {\rm where}  \quad
I_4=\int_0^\infty dy \; y^4 \; f_0(y) \; .
\ee
It is convenient to introduce the dimensionless variable 
\be\label{defga}
\gamma \equiv k \; r_{lin}  \quad {\rm where} \quad   
r_{lin} \equiv \frac{l_{fs}}{\sqrt3} = 
\frac{\sqrt2}{k_{fs}} \; ,
\ee
and consider the transfer function $ T(k) $ as a function of $ \gamma $. $ T(\gamma) $
decreases by an amount of order one for $ \gamma $ increasing by unit. Therefore,
its Fourier transform $ \rho_{lin}(r) $ eq.(\ref{defro}),
decreases with $ r $ having $ r_{lin} $ as characteristic scale. 

\medskip

The dark matter density eq.(\ref{norf}) can be also expressed as an integral over $ y $ 
[eq.(\ref{gil2})] as
\be\label{denDM}
\rho_{DM} = \frac{m}{2 \, \pi^2} \; g \; T_d^3 \; I_2 \; .
\ee
The covariant decoupling temperature $ T_d $ can be related to the effective
number of UR degrees of freedom at decoupling $ g_d $ and the photon 
temperature today $ T_{cmb} $ by using entropy conservation (see for example \cite{gb}): 
\be\label{temp}
T_d = \left(\frac2{g_d}\right)^\frac13 \; T_{cmb}  \;
, \quad {\rm where} \quad T_{cmb} = 0.2348 \; {\rm meV} \; .  
\ee
We obtain the amplitude $ \Delta(k,t) $ today by inserting 
eqs. (\ref{Delteq}), (\ref{V}), (\ref{dini}), 
(\ref{fikp}) and (\ref{dasy}) into eq.(\ref{dchi}) for $ t = t_{\rm today} $ with the result:
\be\label{delhoy}
\Delta(k,t_{\rm today}) = \frac{9 \, \pi}{\sqrt2} \; \frac{M_{Pl}^2}{H_0} \; \Omega_M 
\; b_0 \; b_1 \; A \; (1 + z_{eq}) \; | \Delta_0 | \; T(k) \; 
\left(\frac{k}{k_{eq}}\right)^\frac32 \; 
\left(\frac{k}{k_0}\right)^{\frac{n_s-1}2} \; \log\left(c \; \frac{k}{k_{eq}} \right) \; .
\ee
where $ c = 0.11604 $. 
Inserting eq.(\ref{delhoy}) into eq.(\ref{defro}) yields the density profile today,
\bea\label{perf}
&&\rho_{lin}(r) = \frac{27 \; \sqrt2}{5 \; \pi} \; \frac{\Omega_M^2 \; M_{Pl}^2 \; H_0}{\sigma^2_{DM}} 
\; b_0 \; b_1 \; A \; |\Delta_0| \; \left(k_{eq} \; r_{lin}\right)^{\frac32} \; 
\frac{\left(k_0 \; r_{lin}\right)^{\frac{1-n_s}2}}{r} \; 
\int_0^{\infty} d\gamma \; N(\gamma) \; \sin\left(\gamma \, \frac{r}{r_{lin}} \right) \; 
 , \cr \cr
&& r_{lin} \; \rho_{lin}(0) =  \frac{27 \; \sqrt2}{5 \; \pi} \; \frac{\Omega_M^2 \; 
M_{Pl}^2 \; H_0}{\sigma^2_{DM}} \; b_0 \; b_1 \; A \; |\Delta_0| \; \left(k_{eq} \; 
r_{lin}\right)^{\frac32} \;
\left(k_0 \; r_{lin}\right)^{\frac{1-n_s}2} \; \int_0^{\infty} d\gamma \; \gamma \; N(\gamma) \; .
\eea
where
$$
N(\gamma) \equiv \gamma^{n_s/2-1} \; \log\left(\frac{c \; 
\gamma}{k_{eq} \; r_{lin}}\right) \; T(\gamma) \; .
$$
Notice that there are no free parameters here. All parameters here are known
cosmological parameters and the parameter $ Z $ determined by eq.(\ref{Z}).

From these results we compute and analyze the surface gravity and the density profile
in the sections below.

\section{Properties of the linear density profile and the surface gravity}\label{pld}

It is very useful to relate the free streaming length to the phase-space density $ \rho/\sigma^3 $
\citep{dep,dep2,dm1,dm2}. $ \rho/\sigma^3 $ is invariant under the cosmological 
expansion and decreases due to gravitational clustering (self-gravity interactions). 
The phase-space density before structure formation 
($ \rho_{DM}/\sigma^3_{DM} $) and today can be related as \citep{dm2}
\be\label{Z}
\frac{\rho_s}{\sigma^3_s} = \frac1{Z} \; \frac{\rho_{DM}}{\sigma^3_{DM}} \; .
\ee
where $ \rho_{DM}/\sigma^3_{DM} $ is the constant phase-space density before
the MD era. The constant phase-space density today 
\be\label{gil}
\frac{\rho_s}{\sigma^3_s} \sim 5\times 10^3 ~
\frac{\mathrm{keV}/\mathrm{cm}^3}{\left( \mathrm{km}/\mathrm{s}
\right)^3} = (0.18 \;  \mathrm{keV})^4 \; ,
\ee
follows from dSphs observations \citep{gilmore}. 
The range of values of the factor $ Z $ is discussed below and in sec. \ref{conclu}.

We obtain the primordial DM dispersion velocity $ \sigma_{DM} $ from eqs. (\ref{norf}), 
(\ref{roc}) and (\ref{Z}) \citep{dm2},
\be\label{sig2}
\sigma_{DM} = \left(3 \, \; M_{Pl}^2 \; H_0^2 \; \Omega_{DM} \; \frac1{Z} 
\; \frac{\sigma^3_s}{\rho_s} 
\right)^{\frac13}
\ee
This expression is valid for {\bf any kind} of DM particles.
We find using eq.(\ref{lfs}) for the free streaming length,
eq.(\ref{sig2}) for $ \sigma_{DM} $, and eq.(\ref{gil}),
\be\label{r0Z}
r_{lin}  = \frac{l_{fs}}{\sqrt3} 
= \frac{207.6}{Z^\frac13} \; \; {\rm kpc} 
= 96.37 \; \left(\frac{10}{Z}\right)^{\frac13} \; \; {\rm kpc} \quad
{\rm and} \quad \frac1{\sigma^2_{DM}} = 2.358 \; 10^{13} \; Z^\frac23 \; .
\ee
The velocity dispersion $ \sigma_{DM} \sim 10^{-7}  \; Z^{-\frac13} <  10^{-7} $ 
is very small since it does not take into account the self-gravity contrary to 
$ \sigma_s \sim 10^{-5} $. $ \sigma_{DM} $ is just the covariant primordial 
velocity dispersion.

The linearized Boltzmann-Vlasov equation with the given initial conditions
eqs.(\ref{Delteq})-(\ref{dini}) provides a single solution that can be considered a galaxy
configuration with characteristic size given by the linear scale $ r_{lin} $ which
is of the order of the free-streaming length eq.(\ref{defga}) 
$ r_{lin} \sim l_{fs} $. The length $ r_{lin} $ approaches the halo
radius eq.(\ref{datos}) for the largest galaxies, $ r_{lin} \gtrsim  r_0 $. 
Taking as initial conditions eq.(\ref{Delteq}) multiplied by a unit random gaussian field 
plus taking into account non-linear effects would give a variety of galaxy 
configurations with smaller masses and sizes.

\medskip

Inserting eq.(\ref{r0Z}) $ r_{lin} $ and $ 1/{\sigma^2_{DM}} $, 
and the values eq.(\ref{ns}) in eq.(\ref{perf}) yields for the density 
profile and the surface gravity:
\bea\label{perf2}
\rho_{lin}(r) =  (5.826 \; {\rm Mev})^3 \; 
\frac{Z^{n_s/6}}{r} \; 
\int_0^{\infty} d\gamma \; N(\gamma) \; \sin\left(\gamma \, \frac{r}{r_{lin}} \right)
&& \\ \cr
\mu_{0 D} = r_{lin} \; \rho_{lin}(0) = (5.826 \; {\rm Mev})^3 \; 
Z^{n_s/6} \;
\int_0^{\infty}  d\gamma \; \gamma \; N(\gamma) && \label{perf3}
\eea
where $ n_s/2 -1 = -0.518, \; n_s/2 = 0.482, \; n_s/6 =0.160 $,
$$
N(\gamma) \equiv \gamma^{n_s/2-1} \; \log\left({\widehat c} \;  
Z^{\frac13} \; \gamma\right) \; T(\gamma) \; ,
$$
and 
$ {\widehat c} = 43.6 $.

\begin{figure}[h]
\begin{turn}{-90}
\psfrag{"dengifd.dat"}{Fermi-Dirac}
\psfrag{"dengibe.dat"}{Bose-Einstein}
\psfrag{"r2dengimb.dat"}{Maxwell-Boltzmann}
\includegraphics[height=17cm,width=10cm]{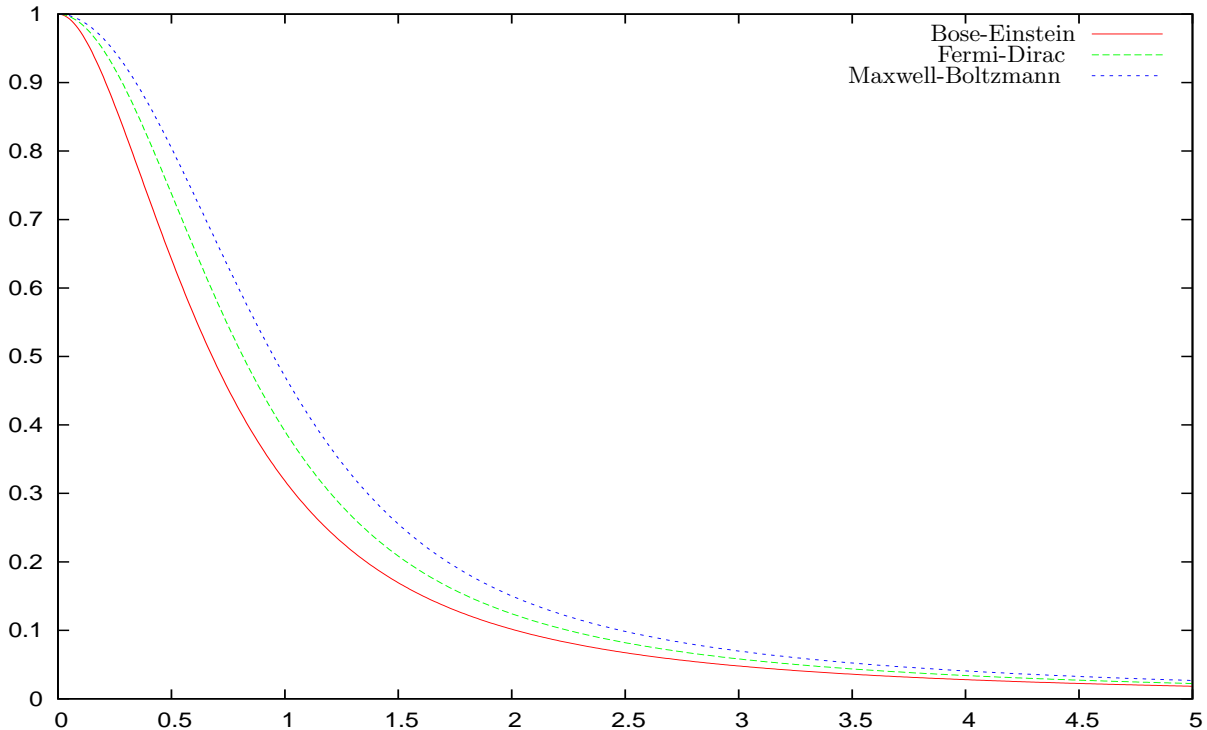}
\end{turn}
\caption{The profiles $ \rho_{lin}(r)/ \rho_{lin}(0) $ vs. $ x $, where $ x \equiv r/r_{lin} $
for Fermions and Bosons decoupling ultrarelativistically and for particles decoupling 
non-relativistically (Maxwell-Boltzmann statistics). The bosons profile is the more peaked,
the MB profile is the shallowest and the fermions profile is lying in-between.}
\label{3perf}
\end{figure}

\begin{table}[h]
\begin{tabular}{ccc} \hline  
Particle Statistics & $ \mu_{0 D} = r_{lin} \, \rho_{lin}(0) $   & 
$ r_{lin}^2 \;  \rho_{lin}''(0)/\rho_{lin}(0) $ \\
\hline \hline 
  Bose-Einstein &  $ (16.71 \; {\rm Mev})^3 \; (Z/10)^{0.16} $ & $ -5.50 $ \\
\hline 
  Fermi-Dirac & $ (15.65 \; {\rm Mev})^3 \; (Z/10)^{0.16} $ & $ -2.74 $ \\
\hline 
  Maxwell-Boltzmann &  $ (14.73 \; {\rm Mev})^3 \; (Z/10)^{0.16} $ & $ -1.83 $\\
\hline   
\end{tabular}
\caption{Values obtained of the surface gravity $ \mu_{0 D} = r_{lin} \, \rho_{lin}(0) $ 
for Fermions and Bosons decoupling ultrarelativistically 
and for particles decoupling non-relativistically (Maxwell-Boltzmann statistics).
[The exponent of $ Z $ originates in the primordial power $ n_s/6 = 0.16 $].
The comparison of these theoretical results  for 
$ \mu_{0 D} = r_{lin} \, \rho_{lin}(0) $ with the observational 
value eq.(\ref{valmu}) indicates that $ Z \sim 10-100 $ and therefore that the DM particle mass is in 
the keV range [see eq.(\ref{mgdeq})]. In any case, the 
agreement between the linear theory and the observations is 
{\bf remarkable}.}
\end{table}

We plot in fig. \ref{3perf} the ratio
\be\label{perfun}
\frac{\rho_{lin}(r)}{\rho_{lin}(0)} \equiv \Psi(x) =
\frac{\int_0^{\infty} N(\gamma) \; \sin\left(\gamma \, x \right) \; 
d\gamma}{x \; \int_0^{\infty} \; \gamma \; N(\gamma) \; d\gamma} \; , \;
x \equiv r/r_{lin} \; ,
\ee
for Fermions (FD) and Bosons (BE) decoupling 
ultrarelativistically and for particles decoupling non-relativistically 
[Maxwell-Boltzmann statistics (MB)]. Notice that $ \Psi(x) $ 
is {\bf independent } of the length scale $ r_{lin} . \;  \Psi(x) $ only depends
on the cosmological parameters with a mild logarithmic dependence on $ Z $, 
as shown by eqs.(\ref{perf2})-(\ref{perf3}).

The theoretical results for $ \mu_{0  D} $ displayed in Table I confronted to
the observed value eq.(\ref{valmu}) suggest the values $ Z \sim 10-100 $. 
We choose for the plots a typical value $ Z = 10 $.
However, the same picture is obtained for all
$ 1 < Z < 10^4 $ since the dependence on $ Z $ is mild.
This is consistent with the independent analysis on the range of $ Z $ in \cite{dm2}.

\medskip

The displayed profiles are clearly {\bf cored}, as expected, since $ T(k) $ decays for 
$ k > k_{fs} \sim 1/r_{lin} $. 
Moreover, the profile eq.(\ref{perf2}) is flat at  $ r = 0 $
with a negative concavity there, namely $ \rho_{lin}'(0) = 0 $ and 
$ \rho_{lin}''(0) < 0 $. More explicity,
$$
\frac{\rho_{lin}(r)}{\rho_{lin}(0)} \buildrel{r \ll r_{lin}}\over= 
1 + \frac{r^2}2 \; 
\frac{\rho_{lin}''(0)}{\rho_{lin}(0)} +{\cal O}(r^4)= 1 - \frac{x^2}6 \; 
\frac{\int_0^{\infty} \gamma^{2.482} \; \log\left({\widehat c} \;  
Z^{\frac13} \; \gamma \right) \; 
T(\gamma) \; d\gamma}{\int_0^{\infty} \gamma^{0.482} \; 
\log\left({\widehat c} \;  Z^{\frac13} \; \gamma \right) \; 
T(\gamma) \; d\gamma} +{\cal O}(x^4) \; .
$$
We display in Table I the values of $ r_{lin} \, \rho_{lin}(0) $ and 
$ r_{lin}^2 \;  \rho_{lin}''(0)/\rho_{lin}(0) $ 
for the three particle statistics: FD, BE and MB. We find that 
$ \rho_{lin}(0)_{BE} > \rho_{lin}(0)_{FD} > \rho_{lin}(0)_{MB} $. 
We display in fig. \ref{3perf} the profiles $ \rho_{lin}(r)/ \rho_{lin}(0) $
as functions of  $ x = r/r_{lin} $.
The more peaked density profile is the one for bosons (BE) and the more shallow is the 
non-relativistic one (MB). The fermions profile being in-between the two other profiles.

\section{Concluding remarks and the DM particle mass in the keV scale}
\label{conclu}

The astronomical observations tells us that the value of the surface gravity 
$ \mu_{0 D} = r_0 \; \rho(0) $ is {\bf universal}. 
Therefore, we can compute $ \mu_{0 D} $ in the limiting case where the linearized 
Boltzmann-Vlasov equation holds. This is why we {\bf identify} 
$ r_{lin} \, \rho_{lin}(0) $ computed for a spherically symmetric solution
of the linearized Boltzmann-Vlasov equation for self-gravitating DM
with the {\bf observed} value eqs.(\ref{valmu})-(\ref{datos}). 
One representative solution should be enough to obtain the value of the surface 
gravity but a more general treatment for non-spherically symmetrical 
solutions of the {\bf non-linear} Boltzmann-Vlasov equation and/or $N$-body simulations
(and including also baryonic matter) will be necessary to 
prove the universality of $ r_0 \, \rho(0) $. 

\medskip

We can estimate the mass of the galaxies obtained in the linear 
approximation from eqs.(\ref{r0Z})-(\ref{perf2}) as
\be\label{mgal}
M \sim  r^3_{lin} \, \rho_{lin}(0) = 1.8 \; 10^{14} \; M_{\odot} 
\left(\frac{10}{Z}\right)^{\frac{4-n_s}6} \; , \; \frac{4-n_s}6 \simeq 0.506 \; .
\ee
We obtain mass values in the upper range of the observations, as expected.

\medskip

Notice the scaling of the linear profile $ \rho_{lin}(r) $ eq.(\ref{perf2})
obtained here with the primordial spectral index $ n_s $:  $ \rho_{lin}(r) $
decreases as 
$$ 
r^{-1-n_s/2} = r^{-1.482}  \quad {\rm for}  \quad  r \gg r_{lin} \; .
$$
The value of this
exponent is in agreement with the universal empirical behaviour recently put forward 
from observations in \citet{wal} and from $\Lambda$CDM simulations in \citet{vass}:
$ r^{-1.6\pm 0.4} $. 
For larger scales we would expect that the contribution from small $ k $ 
modes where nonlinear effects are dominant will give the customary 
$ r^{-3} $ tail.

\medskip

The range of values of $ Z $ 
from analytic approximate formulas both for linear fluctuations and the (non-linear) 
spherical model \citep{dm2} and from $N$-body simulations results 
\citep{numQ,numQ2,numQ3,numQ4,numQ5} is given by
$$ 
1 < Z < 10000 \; .
$$
We find that the surface gravity computed from the linearized Boltzmann-Vlasov equation
reproduces very well the observed value of the energy scale eq.(\ref{valmu})
for the three different particle statistics provided $ Z \sim 10 - 100 $ for dSphs. 
Nonlinear effects should improve the theoretical values of 
the surface gravity $ \mu_{0 D} = r_{lin} \, \rho_{lin}(0) $ in 
Table I including the contributions from 
large scales (small $k$ modes). Notice from eq.(\ref{perf3}) that the theoretically
computed $ \rho_{lin}(r) $ and $ \mu_{0 D} $ have a {\bf mild} 
dependence on $ Z $, the only parameter here which is not known with precision.

Anyhow, the agreement between the linear theory and the observations is already {\bf remarkable}.
The comparison of our theoretical values for $ \mu_{0 D} $ displayed in 
Table I and the observational value eq.(\ref{valmu}) indicates that 
$ Z \sim 10-100 $ for dSphs. 

Notice that $ r_{lin} $  in eq.(\ref{r0Z}) decreases with $ Z $ as 
$ Z^{-\frac13} $, while $ \rho_{lin}(0) $  in eq.(\ref{perf3}) grows with $ Z $ as 
$ Z^{(n_s+2)/6} \; \ln Z = Z^{0.493} \; \ln Z $. 

\medskip

From Table I and eq.(\ref{r0Z}) we obtain for the density contrast between the 
galaxy center and the average DM density
$$
\frac{\rho_{lin}(0)}{\rho_{DM}} \simeq 2 \times 10^4 \; \left(\frac{Z}{10}\right)^{\frac{n_s+2}6}
$$
for FD particles and similar results for the BE and MB statistics. 
The value obtained here is smaller by about
a factor ten than observations \citep{SP}.

\medskip

In summary, the solution of the linearized Boltzmann-Vlasov equation presented here provides
an analytic and explicit approximative picture of a galaxy. 
Although nonlinear effects and baryons are
not taken into account, this simple description qualitatively reproduces the main
characteristics of a galaxy. Moreover, the agreement is even approximatively quantitative for
$ r_{lin} $ eq.(\ref{r0Z}) with the observed halo radius. Similarly for $ M $  eq.(\ref{mgal}) 
with the observed galaxy mass in the limiting case of large size galaxies. 

\medskip

Combining eqs.(\ref{sigma}), (\ref{denDM}), (\ref{temp}) and (\ref{Z}) we can express
$ m $ and $ g_d $ as
\bea\label{msola}
 m^4 &=& \frac{2 \; \pi^2}{3 \; \sqrt3} \; \frac{Z}{g} \; 
\frac{\rho_s}{\sigma^3_s} \; \frac{I_4^{\frac32}}{I_2^{\frac52}}  \quad , \quad
 m = 0.2504 \; \left(\frac{Z}{g}\right)^\frac14 \; \; 
\frac{I_4^{\frac38}}{I_2^{\frac58}} \; \mathrm{keV}  \; ,  \\ \cr
 g_d &=& \frac{2^\frac14}{3^\frac{11}8 \; \pi^\frac32} \; 
\frac{g^{\frac34}}{\Omega_{DM}} \; \frac{T_{cmb}^3}{M_{Pl}^2 \; H_0^2} \; 
\left(\frac{Z \; \rho_s}{\sigma^3_s}\right)^\frac14 
\left(I_2 \; I_4 \right)^{\frac38} = 35.96 \; Z^\frac14 \; g^\frac34 \; 
\left(I_2 \; I_4 \right)^{\frac38} \label{gdm} \; .
\eea
For example, for fermions and bosons that decouple ultrarelativistically
at thermal equilibrium eqs.(\ref{msola}) and (\ref{gdm}) yield  \citep{dm2}
\be\label{mgdeq}
 m = \left(\frac{Z}{g}\right)^\frac14 \; \mathrm{keV} \; \times
\left\{\begin{array}{l}
         0.568~~~\mathrm{Fermions} \\
              0.484~~~\mathrm{Bosons}      \end{array} \right. \quad , \quad
g_d = g^\frac34 \; Z^\frac14 \; \times \left\{\begin{array}{l}
         155~~~\mathrm{Fermions} \\
              180~~~\mathrm{Bosons}      \end{array} \right. \; . 
\ee
Notice that $ 1 < Z^\frac14 < 10 $ for $ 1 < Z < 10000 $.

The range of values $ 1 < Z < 100 $ discussed above and
eqs.(\ref{msola}) and (\ref{mgdeq}) imply that the DM particle mass
is in the keV range.

The DM particle mass 
$ m $ grows as $ Z^{\frac14} $ according to eq.(\ref{msola}) (or as $ Z^{\frac13} $ for DM 
particles decoupling non-relativistically \citep{dm2}).
For example, wimps at $ m = 100 $ GeV 
and $ T_d = 5 $ GeV \citep{pdg} would require $ Z \sim 10 ^{24} $ \citep{dm2}. 
For $ Z \sim 10 ^{24} $ we find the characteristic scale $ r_{lin} $ eq.(\ref{r0Z})
$$
r_{lin} \sim 0.00208 \; {\rm pc} \sim 438 \; {\rm AU}
$$
For such {\bf small} $ r_{lin} $ the linear profile $ \rho_{lin}(r) $ would appear as a {\bf cusped}
profile when observed at scales of the kpc or larger as eq.(\ref{perfun}) and fig. \ref{3perf} 
show. Cusped profiles are thus clearly associated to heavy DM particles with a huge mass 
$ m $ well above the physical keV scale. Wimps with $ Z \sim 10 ^{24} $ are in
contradiction with the observed value eq.(\ref{valmu}) of the surface gravity 
as shown by Table I.

Independent further evidence for the DM particle mass in the keV scale
were recently given in \citet{tikho,tikho2}. (See also \citet{gilmore}).
DM particles with mass in the keV scale can alleviate CDM problems as the satellite problem \citep{sld}
and the voids problem \citep{wp}. 

The DM particle mass in the keV scale explain
why DM particles were not found in detectors sensitive to particles heavier than $ \sim 1 $ GeV
\citep{CDMS}. In addition, astrophysical mechanisms that can explain the 
$ e^+ $ and $ \bar p $ excess in cosmic rays without requiring DM particles
in the GeV scale or above were put forward in \citet{BBS,Bla,serp}.

\medskip

Our present results for the surface gravity and the density profile,
besides their intrinsic interest giving clues to explain the universal
value of the surface gravity, provide further evidence for the mass 
scale of the DM particle being in the keV scale.

\begin{acknowledgments}
We thank Claudio Destri and Paolo Salucci for useful discussions.
\end{acknowledgments}

\end{document}